\documentstyle[12pt,epsf,epsfig]{article}
\makeatletter						
\def\anti#1{\mathchoice{\@anti{\displaystyle}{#1}}
                       {\@anti{\textstyle}{#1}}%
                       {\@anti{\scriptstyle}{#1}}%
                       {\@anti{\scriptscriptstyle}{#1}}#1}
\def\@anti#1#2{\setbox0\vbox{$#1#2$}
  \rlap{$#1\kern0.25\ht0\overline{\kern-0.25\ht0\phantom{#2}}$\hss}}
\makeatother
\begin{document}
\begingroup
\def\thefootnote{\fnsymbol{footnote}}%
\makeatletter
\def\@makefnmark{\hbox to\z@{$\m@th^{\@thefnmark}$\hss}}%
\makeatother
\begin{center}
\vbox to 0pt{\vss 
\begin{flushright}
EPTCO-96-001	\\
hep-ph/9611348	\\[10mm] 
\end{flushright}}
{\large\bf Symposium Summary\footnote
{Presented at SPIN '96---The 12th.\ International Symposium on High-Energy 
 Spin Physics (Amsterdam, Sept.\,1996).}}
\\[5mm]
Philip G.~Ratcliffe
\\[3mm]
{\small\em 
Istituto di Scienze, Universit\`a di Como, via Lucini 3, 22100 Como, Italy\\
and Istituto Nazionale di Fisica Nucleare -- Sezione di Milano}
\end{center}
\endgroup
\setcounter{footnote}{0}%

\begin{center}
ABSTRACT\\[3mm]
\begin{minipage}{130mm} \small
The plenary presentations of the conference are summarised, highlighting some
aspects that were of particular interest and attempting to link a few of the
topics covered.
Particular emphasis is placed on the problem of deep-inelastic scattering and
questions still to be answered with regard to the distribution functions,
strange-quark and gluon contributions, and the possible r\^ole of orbital
angular momentum.
Passing reference is also made to some of the parallel session presentations.
\end{minipage}
\end{center}

\vspace{3mm}\noindent
{\bf Introduction}\\*
Let me begin by apologising for the only scant coverage that I can make of all
the numerous and interesting parallel presentations; the human
lack of ubiquity being what it is, I could attend only one of any of the
five contemporary sessions.
Thus, in order to be fair on those speakers I did not hear, I shall make
no direct references to the parallel talks and limit myself to oblique
citation only.
I should also point out that I am not an experimentalist and am therefore
not qualified to discuss detector details {\em etc}.
On the other hand, nor shall I present any of the particularly formal
theoretical aspects.
These may, of course, be found in the write-ups of the originals elsewhere in
this volume.

As usual in this series of symposia, the parallel-session presentations ranged
from accelerator physics to polarised sources and from detectors to
phenomenology and theory, with a good deal of overlap.
The physics issues involved cover the full extent of the Standard Model,
including both electro-weak and strong-interaction theory.
However, the large amount of effort being spent on the problem of polarised
deep-inelastic scattering and the related structure functions was particularly
noticeable: approximately a third of the talks, both plenary and parallel, had
to do specifically with this topic and an even larger part was indirectly
connected.
As my own interests lie very much in the this area, the same bias will be amply
reflected here; I apologise to those whose preferences lie elsewhere and refer
them to the other contributions reported in these proceedings.

For the sake of brevity and space, I shall only provide full references to
works not reported herein; talks presented at the conference will be simply
referred to by the authors' names.
In the same spirit, I shall also refrain from the repetition of formul\ae\ and
listing of experimental data.

\vspace{5mm}\pagebreak\noindent
{\bf Polarised Structure Functions}\\*
{\em A Very Brief History}\\*
A rather complete account of the theoretical basis for discussing the nucleon
spin structure can be found in\,[1].
The history of this subject can be said to start in 1966 with the pioneering
work of Bjorken\,[2]:
although not based on the modern theory of Quantum Chromodynamics or even
the Quark-Parton Model, this paper nevertheless set up a framework for
phenomenological sum rules based on current algebra.
Thus, a benchmark for experimental measurement was erected, together with a
starting point for theoretical predictions based on the more sophisticated
approach of perturbative {\small QCD} (for details see J.\,Ellis and S.\,Forte
in this volume).
The historical value of the Bjorken sum rule is encapsulated in Feynman's now
celebrated statement that
\begin{quote} \em \noindent
\llap{``\,}\dots\ its verification, or failure, would have a most decisive
effect on the direction of future high-energy physics.''
\end{quote}

As it stands, the BJ sum rule requires the measurement of both the proton and
neutron spin structure functions.
Thus, in order to render it more experimentally accessible, in 1974 two young
researchers took the liberty of a simplifying assumption to cut it in half and
obtain sum rules for the proton and neutron separately\,[3]; after
considering various possibilities, the nomenclature EJ was finally settled
upon:\footnote
{Reproduction of a notebook fragment of uncertain origins ({\em c.}\,1974) now
preserved in the {\small SLAC} historical archives---unfortunately, the
authors' names have long since been rendered illegible.}
\begin{center}
\epsfig{figure=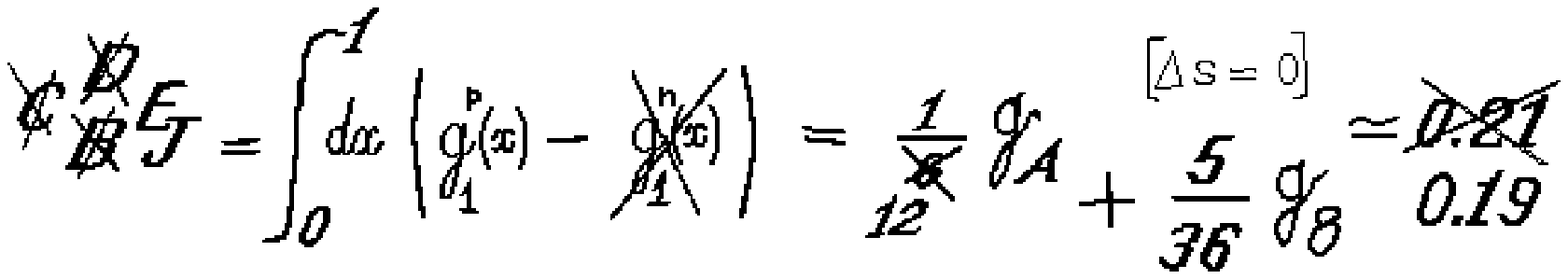,width=13cm}
\end{center}
Note that the {\small PQCD} corrections are an order 10\% effect at typical
experimental energies.
Thus, it was indeed a great surprise to most when in 1988 the {\small EMC}
experiment announced a value\,[4] that was little more than half the
original EJ prediction.
However, with hindsight, one might claim that there was nothing cast-iron about
the assumption made and so the deviation led to the following somewhat
sacrilegious paraphrasing of Feynman's words:\footnote
{Anonymous graffiti found on numerous blackboards.}
\begin{quote} \noindent \em 
\llap{``\,}\dots\ {\em the\/} verification, or failure {\em of the EJ sum
rule}, would have a most {\em in\/}decisive effect on the direction of future
high-energy physics.''
\end{quote}

The questions raised by this discovery were manifold, ranging in gravity from
the reliability of the low-energy $F$ and $D$ parameters to a possible failure
of {\small PQCD}; certainly the EJ assumption was thrown into doubt but equally
the possibilities arose of, {\em e.g.}, orbital angular momentum contributions,
and large gluon or strange-quark polarisations.
All these questions have stimulated research that has undoubtedly led to a
deeper insight into the structure of the proton and to a greater interest in
the subject experimentally.
In any case, it is now perhaps time to try and set the records straight:
\begin{quote} \noindent \em 
\llap{``\,}{\em The very existence of the BJ sum rule has \underbar{already}
had\/} a most decisive effect on the \dots''
\end{quote}
while
\begin{quote} \noindent \em 
\llap{``\,}{\em The failure of the infamous EJ sum rule has \underline{also}
had\/} a most decisive effect on the \dots''
\end{quote}
The former as a spur to investigating the field and the latter as witnessed by
the tremendous effort that has been made in recent years to understand this
problem.

\vspace{5mm}\noindent
{\em Hyperon Semi-Leptonic Decays}\\*
At this point I shall avail myself of the summary speaker's prerogative and
take the opportunity to touch briefly upon a problem not explicitly discussed
during the conference but that is nevertheless of some importance in the
analysis of nucleon spin sum rules: namely, the input coming from the analysis
of the hyperon semi-leptonic decays. 

While the BJ sum rule only requires knowledge of the neutron $\beta$-decay
constant ($g_A/g_V=F+D$), the EJ sum rule for either proton or neutron
necessitates the input of both the SU(3) parameters $F$ and $D$ separately.
The various axial couplings responsible for the decays of other octet
baryon states, depending on different linear combinations, allow just
that, {\em providing one knows how to handle the problem of SU(3) violations}.
Most authors either simply accept one of the ``standard'' published sets
of values (more-or-less tacitly) and thus make the usual deduction that the
quarks carry little of the proton spin, or via often {\em ad hoc\/} and almost
always highly model-dependent analyses find very different values that can
then accommodate the {\small EMC} (and subsequent) results at no further
expense.

Since the deductions made as to, {\em e.g.}, the strange quark polarisation can
vary considerably, it is worth while stressing here that the only {\em
completely\/} consistent analysis performed (based on purely relativistic
corrections) on {\em all\/} of the most up-to-date data produces a value\,[5]:
\[ F/D=0.57\pm0.01, \]
with a more than acceptable goodness of fit and agreement with other
evaluations of related Standard Model parameters.
Thus, it is now clear that, with present-day precision, relativistic
corrections are both necessary and sufficient.
That said, and especially in view of the ever-increasing experimental precision
with which the sum rules are measured (see G.\,Mallot's talk), there is still
plenty of room for improvement, both experimentally and theoretically.

\vspace{5mm}\noindent
{\em Parton Orbital Angular Momentum}\\*
Returning to issues concerning the proton spin structure more directly: a
contribution that rightly aroused great interest, as a new handle on this
problem, was that by X.\,Ji on the inclusion of orbital angular momentum into
the parton picture of the nucleon and its possible experimental detection.
The problem is an old one although, until now, almost untouched in the standard
formalism based on the parton model and {\small PQCD}.
On the other hand, if the so-called ``gluon anomaly'' explanation of the EJ sum
rule failure contains something of the truth, then there must be a large
orbital angular-momentum contribution, compensating that of the gluon helicity.
Thus, it is natural to ask how such a contribution might be dealt with formally
and whether or not it could have any, more direct, phenomenological
consequences.

In relation to the problem of $Q^2$ evolution (\`a la {\small GLAP}), it has
been known for some time that at some energy scale a large orbital
angular-momentum component will be generated radiatively\,[6].
X.\,Ji has now completely formalised this observation by defining all the
relevant quark and gluon operators and by relating their matrix elements to
those of the energy-momentum tensor.
He then suggests that they may be accessible in a process he defines as
``deeply-virtual'' Compton scattering.
Should this be realisable in practice, the door would open into an entirely new
field of parton distribution measurements: namely, orbital angular-momentum
densities.

\vspace{5mm}\noindent
{\em Deep-Inelastic Scattering Data and Analysis}\\*
For those who prefer the concreteness of results already ``under the belt'',
let me now examine the present experimental situation, as illustrated by
G.\,Mallot and discussed from the theoretical point of view by J.\,Ellis,
J.C.\,Collins and S.\,Forte.
The precision with which the nucleon spin structure function $g_1$ is now
measured at the 3--4\% level (taking the BJ sum rule value of $\sim$0.17 as a
benchmark).
As a sign of the subject's maturity, recall that even relatively recently
the goal for the {\em un\/}polarised structure functions was still said to be a
10\% accuracy\,[7].
Indeed, here J.\,Ellis pointed out that the value for $\alpha_s$ extracted
from the BJ sum rule is now highly competitive with those from other sources.
Moreover, S.\,Forte demonstrated that, from the scaling violations alone, one
is also beginning to acquire some sensitivity to the size of the gluon
polarisation in the nucleon, which was quoted as $\Delta{G}=1.0\pm0.4$ (for
$Q^2=1\;$GeV$^2$).

The {\small SMC} ({\small CERN}) and {\small SLAC} groups have both moved on to
measuring other quantities besides $g_1$: {\em e.g.}, $\Delta q$, $g_2$, higher
twist, {\em etc}.
And, of course, we still have more data to look forward to, including those of
the Hermes collaboration at {\small HERA}.
In the case of the transverse spin structure function, $g_2$, one now has the
precision required to observe possible deviations from the Wandzura-Wilczek sum
rule (again based on certain plausible but far-from guaranteed
assumptions\,[8]); and there may already even be some indications in this
direction in the preliminary neutron data.
As echoed by many speakers, J.\,Collins also mentioned the r\^ole that
transverse spin can play, particularly in the quest for understanding
non-perturbative and chiral-symmetry breaking effects.

\vspace{5mm}\noindent
{\em Remaining Questions and Future Directions}\\*
There are several areas that merit further attention and that are, in fact, all
essentially on the agenda:
\begin{description}
\item[a)] \underline{Higher Twist} --
We are now beginning to get a handle on this sticky subject (see the {\small
SLAC} analysis of their low-$Q^2$ data).
The importance of this problem is clearly twofold: firstly, one would like to
have such contributions under control to improve the reliability of the
structure function analyses, and secondly, it is a study of interest in its own
right, where spin, with its intimate links to chiral symmetry, can play an
important r\^ole.
This is an area in which the Hermes collaboration can also provide some clues.
\item[b)] \underline{Low-$x$ Extrapolation} --
Here one should perhaps give credit to Close and Roberts for their insistence
on this point\,[9].
Although perhaps premature to abandon the asymptotic predictions of Regge
theory entirely, one should clearly tread carefully, especially in the light of
the unpolarised {\small HERA} data.
Moreover, the resummation of the $\log{x}$ terms may be rather more subtle in
the polarised case.
The answers to these questions could most obviously be sought at a {\em
polarised\/} {\small HERA} (see the talk by A.\,Sch\"afer).
\item[c)] \underline{Flavour Separation} --
Purely fully inclusive {\small DIS} cannot separate the various flavour
contributions; the valence distributions can be isolated with some degree of
precision but the sea quarks are not easily accessible.
Since it is precisely to the sea quarks ({\em i.e.}, strange quarks) that the
EJ sum rule discrepancy is attributed (whether it be intrinsically or via the
anomaly), a better handle on this sector is of some urgency.
While the {\small SMC} is providing some information in this direction via
semi-inclusive data, a better facility for such study will be provided by
{\small RHIC} (see the talks by Y.\,Makdisi and H.\,En'yo).
There one will be able to measure the distributions separately via the study of
$W^\pm$ production in different kinematic regions.
\item[d)] \underline{Gluon Spin} --
While {\small DIS} scaling violations can give some information on the
magnitude of the gluon polarisation, complete details and precision require the
study of such processes as direct photon production (as will be measured at
{\small RHIC}), which should provide important shape information.
Again, a {\em polarised\/} {\small HERA} could reduce the present error on the
size of $\Delta{G}$ via its larger $Q^2$ lever-arm and polarised proton beam on
a polarised hadronic target, as discussed by A.\,Sch\"afer, could also give
access to the shape.
\end{description}

The developments at {\small RHIC} were discussed in the two talks by
Y.\,Makdisi and H.\,En'yo, who indicated the possibilities of measuring such
quantities as the gluon polarisation and separating the flavour contributions,
and underlined the complementarity to lepton-hadron scattering experiments at
{\small CERN}, {\small SLAC}, {\small HERA} and {\small CEBAF}.
The advantages of this machine are clearly its high luminosity and beam
polarisation, which coupled to the energy range available will provide access
to the medium-$x$ range (where the polarised structure functions are large).
The possibility of spin rotators will also allow measurements of the various
transverse-spin structure functions (including transversity).

\vspace{5mm}\noindent
{\em Low-$x$ Extrapolation}\\*
Before closing this long section on {\small DIS} I shall briefly examine a
problem, already touched upon, that is of interest both theoretically and in
connection with data analysis: namely, the problem of the small-$x$ behaviour.
It is perhaps worth spending a few words to underline the importance of a
correct treatment of this aspect, particularly in view of the important
differences with respect to the unpolarised case.

As evident from the analyses presented by S.\,Forte, theoretical deduction from
experimentally observed behaviour in polarised measurements is far less simple
than one might be led to imagine, based on experience and knowledge of the
unpolarised phenomenology.
At the risk of over-simplifying, the source of the difficulties may be said to
reside in the non-positivity of the spin distributions.
The point is then that while the unpolarised distributions behave essentially
monotonically with $Q^2$ ({\em i.e.}, $F_2$ increases at small $x$ and
decreases at large $x$ for increasing $Q^2$, owing to the Bremsstrahlung
driving a degradation of the momentum distributions), the polarised
distributions may have a much richer behaviour owing to cancellations of the
various positive and negative contributions.

Moreover, whereas $F_2$ is already dominated by the sea ({\em alias\/} the
Pomeron) for $x$ below $\sim$0.1 say, the same is far from being true for
$g_1$, as witnessed by the opposite signs of $g_1^p$ and $g_1^n$ in this
region.
This means, in particular, that extrapolation from the present values of $x$
(even those of the {\small SMC}) will be highly model dependent, the error
introduced depending on the extremity of the limiting behaviour one is willing
(or desirous) to admit.
Moreover, it is not obvious at what values of $x$ (if any) one can reasonably
expect ({\em without\/} the advantage of hindsight) that the extrapolation
should become unambiguous.

At this conference some note was taken of the present trend in the neutron data
in the small-$x$ region, which appear well described with a form of a simple
(albeit large) power:
\[ g_1^n(x)\sim\frac{-0.02}{x^{0.8}}. \]
Such a behaviour, if extrapolated to zero, could more than double the present
estimates for the integral of $g_1^n$.
This is, however, very misleading; a form such as
\[ g_1^n(x)\sim\frac{-0.07}{x^{0.5}}(1-4x), \]
which gives a very good description of the data over a limited region, would
give a contribution much more in line with the standard estimates.

In other words:
cancellations between two relatively positive and negative contributions (one
growing and one falling) can mimic even highly divergent behaviour over any
{\em finite\/} range of $x$.

\vspace{5mm}\noindent
{\bf Single-Spin Asymmetries}\\*
Let me now turn to an age-old problem in hadronic physics: that of the large
(transverse) single-spin effects observed in hadron-hadron scattering, even at
the large energies and the relatively large momentum transfers presently
accessible.
Once more we were treated to K.\,Heller's customary talk on a subject that has
been troubling phenomenologists now for approximately two decades.
There are essentially two distinct experimental situations in which such
phenomena are observed:
\begin{itemize}
\item hyperons produced semi-inclusively in hadron-hadron collisions are
polarised along the normal to the production plane,
\item mesons are produced asymmetrically (left-right) with respect to the plane
defined by the beam-axis and the (transverse) polarisation direction in
semi-inclusive (singly) polarised hadron-hadron scattering.
\end{itemize}

The basic puzzle lies in the fact that descriptions, either in terms of the
``complicated'' interplay of many amplitudes or ``simple'' hard partonic
scattering, naturally lead to negligible effects.
The former owing to cancellations between uncorrelated amplitudes and the
second because the necessary spin-flip and imaginary phase are not generated in
na{\"\i}ve massless {\small PQCD} tree diagrams.
On the other hand these phenomena are rendered all the more interesting by the
regularities displayed; the asymmetries (typically of the order of a few tens
of percent):
\begin{itemize}
\item increase with $x_F$,
\item increase with $p_T$ (up to about 1\,GeV),
\item obey simple SU(6) relations with regard to signs,
\item are approximately energy independent.
\end{itemize}
And while many models are able to explain some of the effects, there are
certain irregularities not all explicable in any given model:
\begin{itemize}
\item ${\cal P}_{\Xi^-}$ increases with energy,
\item ${\cal P}_{\Sigma^+}$ decreases with energy,
\item ${\cal P}_{\Xi^-}$ does {\em not\/} increase with $x_F$,
\item ${\cal P}_{\anti\Xi^-}$ and ${\cal P}_{\anti\Sigma^+}$ are
{\em non}-zero.
\end{itemize}

The attitude of the ``non-spin'' physicist might be summed up as assuming that
such effects will eventually go away for large enough transferred momenta
(where {\small PQCD} should correctly describe the dominant contributions) and
that in any case the phenomenology is too complex to be of interest.
As a counter to this, first of all, note that the regularities described above
are very suggestive of simple mechanisms while the few irregularities and {\em
possible\/} disappearance of polarisation at large $p_T$ only serve to enhance
the importance of understanding such processes.
In particular, the point at which the hard dynamics really takes over could
provide vital information on the strength and r\^ole of the sub-asymptotic
and/or non-perturbative dynamics and, in this specific case, possible
indications as to the r\^ole of chiral symmetry breaking.
Moreover, there is also the eventuality of the effects resisting unabated (as
they still do) even up to typically ``perturbative'' scales, thus posing a
question that could no longer then be simply ignored.

\vspace{5mm}\noindent
{\bf Weak Interactions as Probes of Spin Structure and New Physics}\\*
There are a number of the physics issues that can be addressed via the use of
weak interactions:
these include, besides the obvious Standard Model physics, the search for new
physics and also an alternative approach to hadronic matrix elements related to
those measured in polarised {\small DIS}.

In the search for new physics, and this was exemplified by M.\,Musolf in terms
of a possible new $Z^0$ (or $Z'$) and/or effect four-fermion couplings, the
different helicity structures likely to be exhibited would provide unambiguous
signals.
In this respect, it should always be remembered that at the elementary level
many polarisation asymmetries are naturally of the order of 100\%, especially
in the high-energy limit where masses and thus spin-flip may be neglected.
The fact that they change dramatically (even sign) for new interactions ({\em
viz.} supersymmetry) renders them highly sensitive to the onset of new energy
scales and related physics.

Thus, a modest improvement in the measurement of say $A_{LR}(0^+,0)$ in nuclear
scattering to the level of 1\% would push the lower limits on the mass of a
$Z'$ and the scale of compositeness up to 0.9\,TeV and 16\,TeV respectively.

The so-called ``oblique'' corrections to boson exchange can also lead to
sensitivity to new physics: these are usually parametrised via the $S$ and $T$
parameters.
Low-energy experiments are typically more sensitive to $S$ and the present
level of sensitivity is only just short of revealing possible deviations from
zero.

Measurements involving electro-weak interactions can also provide access to
hadronic currents not accessible with purely electromagnetic probes:
in particular the vector and axial-vector couplings of the $Z$ boson provide a
third combination of {\em up\/}, {\em down\/} and {\em strange\/} currents thus
allowing a measurement of the strange contribution to proton matrix elements.
B.\,Beise, reporting on preliminary results from the {\small MIT}/Bates
experiment {\small SAMPLE}, presented a measurement of the strange magnetic
moment $\mu_s$ for the nucleon.
The first indications appear to be for a small (of a similar magnitude to the
predictions) but positive value;  within errors it is, however, compatible with
zero.
The theoretical estimates for this quantity vary widely; however, they are
generally negative.
There are a number of theoretical approaches to the problem, all having in
common the need to generate strange-quark loops.
This has be done in calculations based on $K$-meson clouds, {\small VMD} poles,
the constituent {\small QPM}, the Skyrme model, bag models, {\em etc}.
It is hoped to take more data in 1997 and so we might look forward to serious
testing of these theoretical models.

\vspace{5mm}\noindent
{\bf Polarised Electrons}\\*
Spin is now becoming a fact of life in storage rings: the natural polarisation,
due to the Ternov-Sokolov effect\,[10], has a maximum theoretical value
of 92.4\% and, in practice, the polarisations obtained now are not far below.
In his account of progress in polarised electron beams, D.\,Barber underlined
that already at {\small HERA} polarised positron beams of around 70\% had been
maintained for up to ten hours and that in the future one can hope for as much
as 80\%.

An example demonstrating the day-to-day importance of spin is the use of
polarised beams at {\small LEP}, where the polarisation is typically between 5
and 10\% (although it should be remembered that 57\% was achieved in 1993).
Using a one gauss-metre radial magnetic field an energy calibration at the MeV
level has been attained ($\Delta{M_Z}\simeq1.5\;$MeV and
$\Delta\Gamma_Z\simeq1.7\;$MeV).

Finally, D.\,Barber was at pains to stress the need for advanced planning:
polarisation cannot be an afterthought; the possibility has to be at least
allowed for at the design stage in order to avoid the risk of its total
preclusion.

Linear colliders have also progressed far in respect of spin physics:
results of the {\small SLD} collaboration at {\small SLAC} were presented by
R.\,Prepost, who particularly emphasised the r\^ole of the machine as a
predominantly polarised facility, with a beam polarisation approaching 80\%.
Measurements of the left-right asymmetry $A_{LR}$ contribute decisively to an
increased precision in $\sin^2\!\theta_W^{\mathrm{eff}}$.
He also discussed the importance of Ring Imaging \v{C}erenkov Detectors in
heavy quark measurements ({\em e.g.}, $A_c$ and $A_b$) and stressed the
importance of polarisation in separating $B$ versus $\anti{B}$ production in
the study of $B$-$\anti{B}$ mixing. 

\vspace{5mm}\noindent
{\bf Medium- and Low-Energy Hadron Spin Physics}\\*
S.\,Vigdor summarised possible tests of symmetry principles accessible in
medium-energy hadron physics, such as T-violation in $\vec{p}-\vec{d}$, which
is under consideration at {\small COSY} in J\"ulich.
This would be the first experiment to use a recently proposed technique to
measure $\Delta\sigma_{\mathrm{tot}}$ (spin) via the spin dependence of the
stored beam lifetime.
Charge-symmetry breaking effects in nuclear forces ({\em i.e.}, non-invariance
under $u\rightleftharpoons{d}$) can also be usefully investigated with the use
of polarisation.
Here the proposed {\small IUCF} search would use polarisation as a filter for
rare processes.

The use of $\vec{p}$ beam (target) and $\vec{d}$ target (beam) will allow
efficient testing of calculations for $\vec{N}+\vec{d}\to{Nnp}$, to address the
question of how three-nucleon forces are manifested in scattering.
The way in which dedicated experiments can address the questions posed in
few-body physics was also discussed by J.\,van den Brand: examples are the
r\^oles of relativistic corrections and three-body forces.

\vspace{5mm}\noindent
{\bf Round Tables}\\*
Besides the numerous parallel sessions, two round-table discussions were
held.
The first, organised by Yu.\,Arestov on the {\small RAMPEX} experiment, a new
spin experiment with 70\,GeV/c protons in Protvino, was aimed at uncovering
possible measurements (particularly of the single-spin type) that would be
suitable for this experiment.
The other, organised by V.\,Hughes on the gluon spin distribution, had the
purpose of concentrating effort towards the search for and measurement of this
important quantity at the various experimental facilities, either presently
operating or proposed for the future.
Many interesting ideas emerged in these discussions and the interested reader
is referred to the reports presented in this volume.

\vspace{5mm}\noindent
{\bf Final Words}\\*
Spin physics has now definitely come of age, as is more than evident from the
wealth of experimental and theoretical results in the field: notably, the
accurate determination of structure functions, electro-weak measurements and
sophisticated calculations within the framework of perturbative {\small QCD}.
There are now a number of experiments dedicated to studying spin properties and
the {\small SLAC} machine may be considered as essentially a polarised
facility.
On the theoretical side, many phenomenologists traditionally engaged in
unpolarised studies have transferred their skills to the spin sector, often
finding a much richer structure allowing a deeper insight into the nature of
particle interactions.
Last but not least, the r\^ole polarisation has as a precision tool should not
be forgotten ({\em e.g.}, the {\small LEP} beam-energy measurement). 

Some of the old problems are still with us and many new ones continue to
appear; as the older physicist who are still with us know and hopefully the
younger ones who are continually appearing will come to appreciate, the field
still clearly has a lot more to offer.

\vspace{20mm}\vfill
{\small\begin{description}
\item{[1]}
M.~Anselmino, A.~Efremov and E.~Leader, {\em Phys. Rep.\/} {\bf 261} (1995)~1.
\item{[2]}
J.D.~Bjorken, {\em Phys. Rev.\/} {\bf 148} (1966)~1467.
\item{[3]}
J.~Ellis and R.L.~Jaffe, {\em Phys. Rev.\/} {\bf D9} (1974)~1444; {\em
  erratum\/} {\em ibid.\/} {\bf D10} (1974) 1669.
\item{[4]}
EMC, J.~Ashman {\em et~al.}, {\em Phys. Lett.\/} {\bf B206} (1988)~364.
\item{[5]}
P.G.~Ratcliffe, {\em Phys. Lett.\/} {\bf B365} (1996)~383.
\item{[6]}
P.G.~Ratcliffe, {\em Phys. Lett.\/} {\bf B192} (1987)~180.
\item{[7]}
S.D.~Ellis, in the proc. of {\em The 1990 Summer Study on High Energy Physics:
  Research Directions for the Decade\/} (Snowmass, Colorado, Jun.--Jul. 1990),
  ed. E.L.~Berger (World Sci., 1992), p.~20.
\item{[8]}
S.~Wandzura and F.~Wilczek, {\em Phys. Lett.\/} {\bf B72} (1977)~195.
\item{[9]}
F.E.~Close and R.G.~Roberts, {\em Phys. Rev. Lett.\/} {\bf 60} (1988)~1471.
\item{[10]}
A.A.~Sokolov and I.M.~Ternov, {\em Dokl. Akad. Nauk. SSSR\/} {\bf 153}
  (1963)~1052.
\end{description}}

\end{document}